\documentclass[twocolumn,aps,prl]{revtex4}

\pdfoutput=1
\usepackage[pdftex]{graphicx}
\usepackage[unicode=true, bookmarks=false,bookmarksnumbered=true,bookmarksopen=true, breaklinks=false,backref=false,pagebackref=false, colorlinks=true]{hyperref}
\hypersetup{pdftitle={ },pdfauthor={ },linkcolor=black,citecolor=black, urlcolor=black, filecolor=black,pdfpagelayout=OneColumn,pdfnewwindow=true,pdfstartview=XYZ, plainpages=false}

\usepackage{color}
\usepackage{amsmath}
\usepackage{braket}
\usepackage[T1]{fontenc}
\usepackage{url}
\usepackage[english]{babel}
\usepackage{dcolumn}
\usepackage{bm}
\usepackage{subfig}
\usepackage{float}
\usepackage{url} 

\newcommand{\Tr}{\mbox{\rm Tr}}
\newcommand{\ReC}{\mbox{\rm Re}}

\newcommand{\be}{\begin{equation}}
\newcommand{\ee}{\end{equation}}
\newcommand{\bea}{\begin{eqnarray}}
\newcommand{\eea}{\end{eqnarray}}
\newcommand{\non}{\nonumber}
\newcommand{\bit}{\begin{itemize}}
\newcommand{\eit}{\end{itemize}}
\newcommand{\mbf}{\mathbf}

\graphicspath{{figures/}}
\usepackage{verbatim}

\begin{document}
\title{Colour Fields Computed in SU(3) Lattice QCD for the Static Tetraquark System}

\author{Nuno Cardoso}
\email{nunocardoso@cftp.ist.utl.pt}
\author{Marco Cardoso}
\email{mjdcc@cftp.ist.utl.pt}
\author{Pedro Bicudo}
\email{bicudo@ist.utl.pt}
\affiliation{CFTP, Departamento de F\'{\i}sica, Instituto Superior T\'{e}cnico, Av. Rovisco Pais, 1049-001 Lisboa, Portugal}

\begin{abstract}
The colour fields created by the static tetraquark system are computed in quenched SU(3) lattice QCD, in a $24^3\times 48$
lattice at $\beta=6.2$
corresponding to a lattice spacing $a=0.07261(85)$ fm.
We find that the tetraquark colour fields are well described by a double-Y, or butterfly, shaped flux tube.
The two flux tube junction points are compatible with Fermat points minimizing the total flux tube length.
We also compare the diquark-diantiquark central flux tube profile in the tetraquark with the quark-antiquark fundamental flux tube profile in the meson, and they match, thus showing that the tetraquark flux tubes are composed of fundamental flux tubes. 
\end{abstract}
\maketitle

\begin{figure}[t!]
\begin{centering}
    \subfloat[\label{fig:tq1}]{
\begin{centering}
    \includegraphics[width=4.3cm]{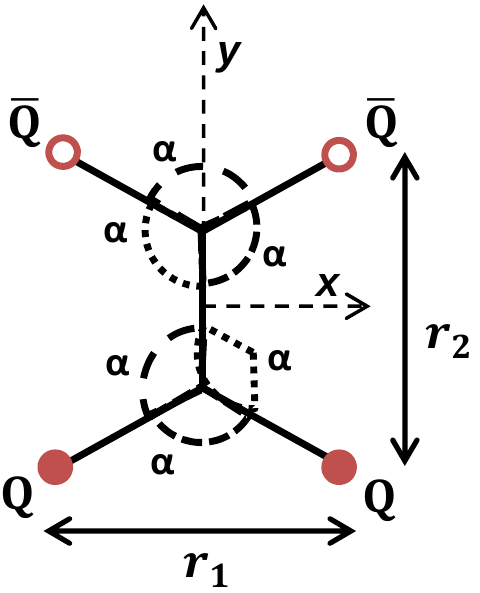}
\par\end{centering}}
    \subfloat[\label{fig:tq0}]{
\begin{centering}
    \includegraphics[width=3.3cm]{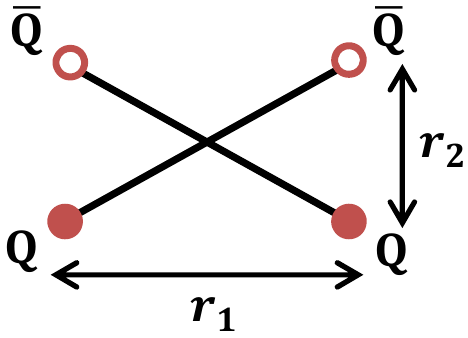}
\par\end{centering}}
\par\end{centering}
    \caption{In the tetraquark flux tube model, the elementary flux tubes meet in two Fermat points, at an angle of  $\alpha=120^{\circ}$  to form a double-Y flux tube, except when this is impossible and the flux tube is X-shaped.}
    \label{fig:tq}
\end{figure}

Multiquark exotic hadrons like the tetraquark and the pentaquark, different from the the ordinary mesons and baryons, have been studied and searched for many years. 
The tetraquark was initially proposed by Jaffe \cite{Jaffe:1976ig} as a bound state formed by two quarks and two antiquarks.
Presently several observed resonances are tetraquark candidates.
Very recently the Belle Collaboration made the tantalizing observation
\cite{Collaboration:2011gja},
in five different $\Upsilon$(5S)  decay channels 
of two new charged bottomonium resonances Z$_b$
with masses of 10610MeV/c$^2$ and 10650MeV/c$^2$ and narrow widths of the order or 15 MeV.  Since all standard bottomonia are neutrally charged, these two new resonances have a  flavour only compatible with $b \, \bar b  \, u  \, \bar d$ tetraquarks.
This is the clearest tetraquark candidate so far observed. Other potential tetraquark candidates have also been observed, however they may still be interpreted differently. 
For instance, in 2003, the X(3872) observed by the Bell Collaboration  \cite{Choi:2003ue,Acosta:2003zx} was suggested as a tetraquark candidate by Maiani et al \cite{Maiani:2004vq}.
In 2004, the D$_{\text{sJ}}$(2632) state seen in Fermilab's SELEX \cite{Jun:2004wn,Cooper:2005zu} was suggested as a possible tetraquark candidate.
In 2009, Fermilab announced the discovery of Y(4140), which may also be a tetraquark \cite{Mahajan:2009pj}.
There are as well indications that the Y(4660) could be a tetraquark state \cite{Cotugno:2009ys}.
The $\Upsilon$(5S)  bottomonium has also been recently suggested to be a tetraquark resonance \cite{Ali:2009es}.
However a better understanding of tetraquarks is necessary to confirm or disprove the  X, Y, Z and possibly also other light resonances candidates as tetraquark states.

\begin{figure}[t!]
\begin{centering}
    \includegraphics[width=6cm]{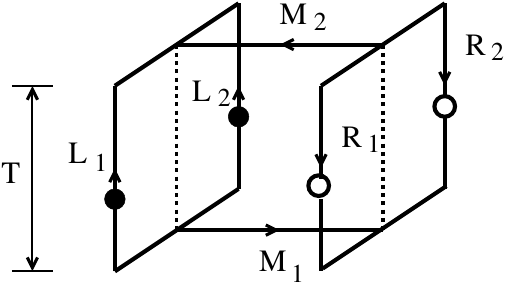}
\par\end{centering}
    \caption{Tetraquark Wilson loop as defined by Alexandrou et al 
\cite{Alexandrou:2004ak}, and by
Okiharu et al \cite{Okiharu:2004ve}.}
    \label{fig:WL_TQ}
\end{figure}

On the theoretical side, the first efforts have been to search for bound states below the strong decay threshold
\cite{Beinker:1995qe,Zouzou:1986qh,Gelman:2002wf,Vijande:2007ix}, 
as it is apparent that the absence of a potential barrier may produce a large decay width to any open channel. 
Recent investigations found that, even above the strong decay threshold, 
the presence of a centrifugal barrier in high angular momentum multiquarks may increase the stability of the system \cite{Karliner:2003dt,Bicudo:2010mv}.

In the last years, the static tetraquark potential has been studied in Lattice QCD computations \cite{Alexandrou:2004ak,Okiharu:2004ve,Bornyakov:2005kn}. 
The authors concluded that when the quark-quark are well separated from the antiquark-antiquark, 
the tetraquark potential is consistent with One Gluon Exchange Coulomb potentials plus a four-body confining potential, 
suggesting the formation of a double-Y flux tube, as in Fig. \ref{fig:tq},
composed of five linear fundamental flux tubes meeting in two Fermat points \cite{Vijande:2007ix,Bicudo:2008yr,Richard:2009jv}.  A Fermat, or Steiner, point is defined as a junction minimizing the total length of strings, where linear individual strings join at $120^{\circ}$ angles.
\begin{figure*}[t!]
\begin{center}
 \subfloat[\label{fig:TQ_EB_ape_hyp_r1_8_r2_14_Act_3D_Sim}]{
\begin{centering}
    \includegraphics[width=8cm]{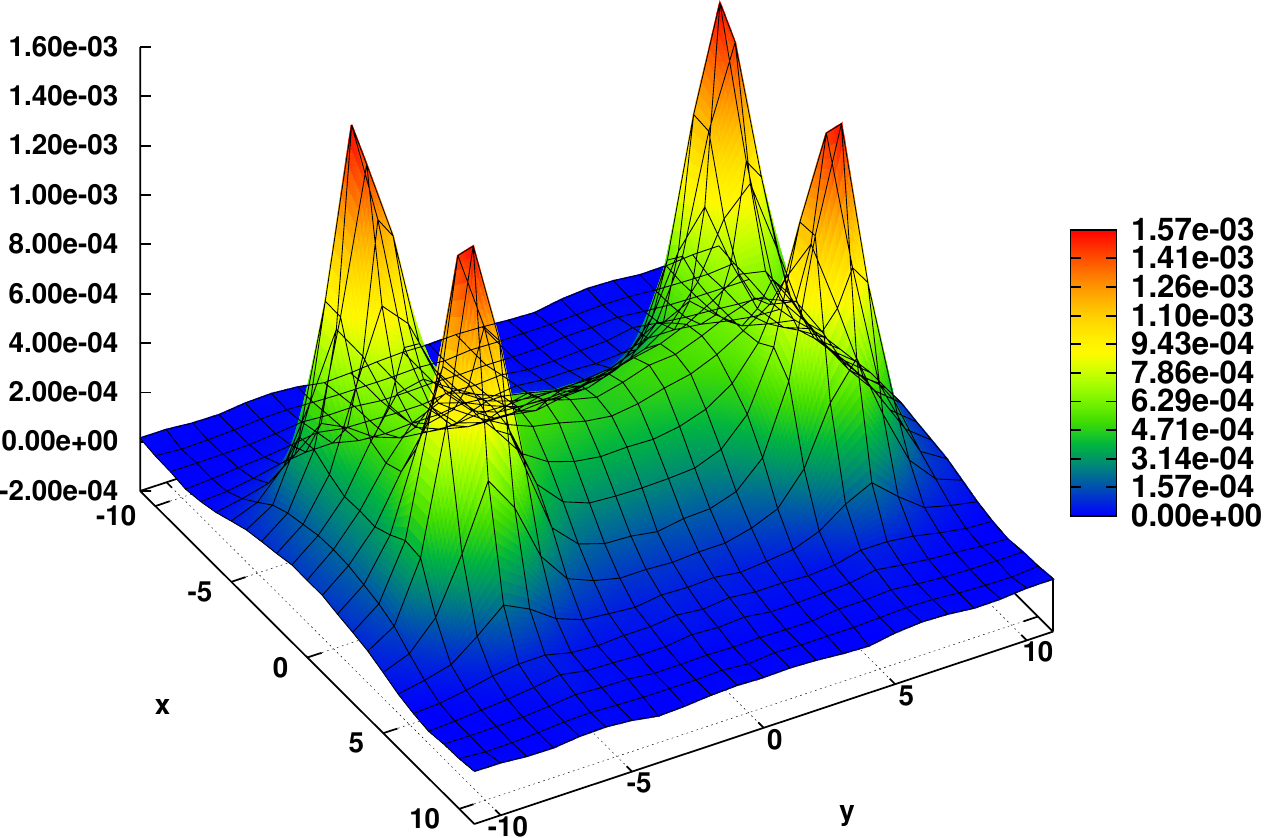}
    \hspace{1cm}
\par\end{centering}}
 \subfloat[\label{fig:TQ_EB_ape_hyp_r1_8_r2_8_Act_3D_Sim}]{
\begin{centering}
    \includegraphics[width=8cm]{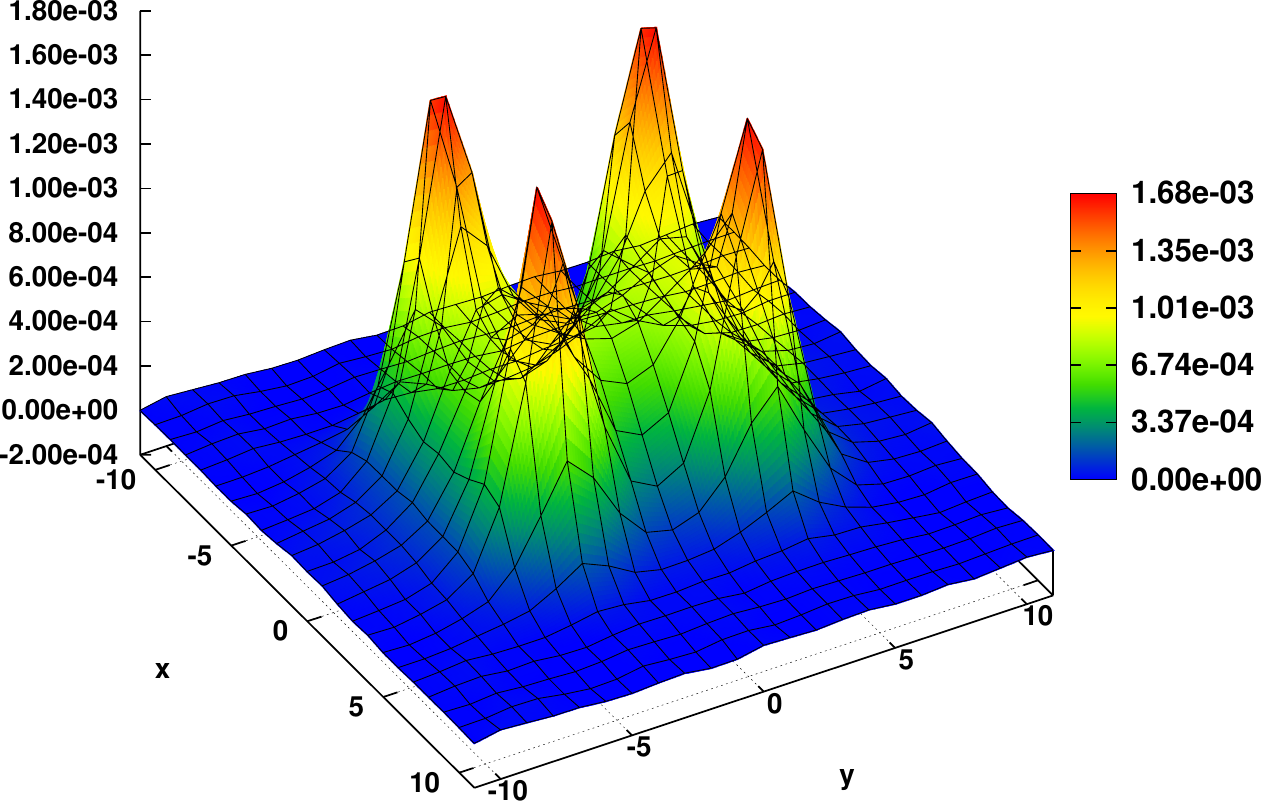}
\par\end{centering}}
\end{center}
    \caption{(a) Lagrangian density 3D plot for $r_1=8, \ r_2=14$. (b) 
We also show the 3D plot for  $r_1=8$ and $r_2=8$,
to illustrate that even at distances where the meson meson dominates the flip-flop potential,
the meson meson mixing with the tetraquark is sufficiently small to produce such a clear a tetraquark double-Y flux tube. 
The results are presented in lattice spacing units (colour online).}
    \label{Act_ape_hyp_r1_8_r2_14}
\end{figure*}
When a quark approaches an antiquark, the minimum potential changes to a 
sum of two quark-antiquark potentials, which indicates a two meson state.
In principle a X-shaped flux-tube as in Fig. \ref{fig:tq0} could also occur,
but the potential minimization always leads in that case to a two-meson potential.
This is consistent with the triple flip-flop potential, minimizing the length, 
with either tetraquark flux tubes or meson-meson flux tubes, 
of thin flux tubes connecting the different quarks or antiquarks 
\cite{Vijande:2007ix,Bicudo:2010mv}.

Here we study the colour fields for the static tetraquark system, with the aim of observing
the tetraquark flux tubes suggested by these static potential computations. 
The study of the colour fields in a tetraquark is important to discriminate between different 
multi-quark Hamiltonian models, 
quark models with two-body interactions only \cite{General:2007bk},
from flip-flop models with a multi-body potential \cite{Bicudo:2010mv}.
Unlike the colour fields of simpler few-body systems, say mesons, baryons and hybrids,
\cite{Ichie:2002dy,Okiharu:2004tg,Cardoso:2009kz,Cardoso:2010kw},
the tetraquark fields have not been previously studied in lattice QCD.

To impose a static tetraquark, we utilize the respective Wilson loop 
\cite{Alexandrou:2004ak,Okiharu:2004ve}
of Fig. \ref{fig:WL_TQ}, 
given by
$W_{4Q} = \frac{1}{3} \Tr \left( M_1 R_{12} M_2 L_{12} \right)$, where
\bea
R_{12}^{aa'} &=& \frac{1}{2}\epsilon^{abc}\epsilon^{a'b'c'}R_1^{bb'}R_2^{cc'}\ , 
\non \\ 
L_{12}^{aa'} &=& \frac{1}{2}\epsilon^{abc}\epsilon^{a'b'c'}L_1^{bb'}L_2^{cc'} \, .
\label{Wilson path}
\eea

The chromoelectric and chromomagnetic fields on the lattice are given by the
Wilson loop and plaquette expectation values,
\bea
\label{fields} 
   \Braket{E^2_i(\mbf r)} &=& \Braket{P(\mbf r)_{0i}}-\frac{\Braket{W(r_1,r_2,T) \,P(\mbf r)_{0i}}}{\Braket{W(r_1,r_2,T)}} 
 \\ \non
    \Braket{B^2_i(\mbf r)} &=& \frac{\Braket{W(r_1,r_2,T)\,P(\mbf r)_{jk}}}{\Braket{W}(r_1,r_2,T)}-\Braket{P(\mbf r)_{jk}} \, ,
\eea
where the $jk$ indices of the plaquette complement the index $i$ of the magnetic field,
and where the plaquette at position $\mbf r=(x,y,z)$ is computed at $t=T/2$, 
\be
P_{\mu\nu}\left(\mbf r \right)=1 - \frac{1}{3} \ReC\,\Tr\left[ U_{\mu}(\mbf r) U_{\nu}(\mbf r+\mu) U_{\mu}^\dagger(\mbf r+\nu) U_{\nu}^\dagger(\mbf r) \right]\ .
\ee
The energy ($\mathcal{H}$) and lagrangian ($\mathcal{L}$) densities are then
computed from the fields, 
\bea
   \Braket{ \mathcal{H}(\mbf r) } &=& \frac{1}{2}\left( \Braket{\mbf E^2(\mbf r)} + \Braket{\mbf B^2(\mbf r)}\right)\ , \\
    \label{energy_density}
   \Braket{ \mathcal{L}(\mbf r) } &=& \frac{1}{2}\left( \Braket{\mbf E^2(\mbf r)} - \Braket{\mbf B^2(\mbf r)}\right)\ .
    \label{lagrangian_density}
\eea

To compute the static field expectation value, we plot the expectation value
    $ \Braket{E^2_i(\mbf r)} $ or   $\Braket{B^2_i(\mbf r)}$ as a function of the temporal extent $T$ of
    the Wilson loop. At sufficiently large $T$, the groundstate corresponding to the
studied quantum numbers dominates, and the expectation value tends to a horizontal plateau. 
In order to improve the signal to noise ratio of the Wilson loop, we use 50 iterations of APE Smearing
with $w = 0.2$ (as in 
\cite{Cardoso:2009kz})
in the spatial directions and one iteration of hypercubic blocking (HYP) in the temporal direction.
 \cite{Hasenfratz:2001hp}, 
with $\alpha_1 = 0.75$, $\alpha_2 = 0.6$ and $\alpha_3 = 0.3$.
Note that these two procedures are only applied to the Wilson Loop, not to the plaquette.
To compute the fields, we fit the horizontal plateaux obtained for each point $\mbf r$ 
determined by the plaquette position, but we consider $z=0$ for simplicity. 
For the distances $r_1$ and $r_2$ considered, we find in the range of $T\in [3,12]$ in lattice units,
horizontal plateaux with a $\chi^2$ /dof $\in [0.3,2.0] $.  
We finally compute the error bars of the fields with the jackknife method.

\begin{figure*}[!htb]
\begin{center}
    \includegraphics[width=4.2cm]{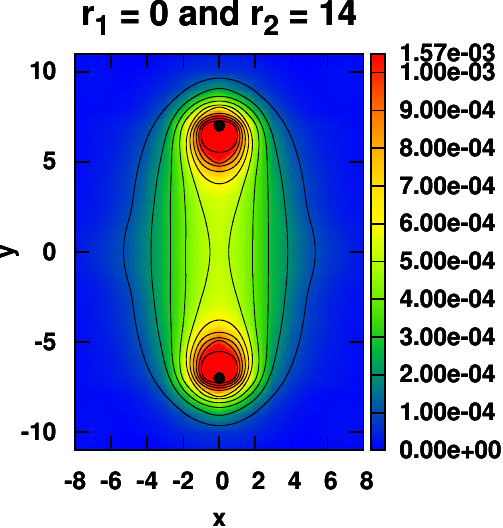}
    \includegraphics[width=4.2cm]{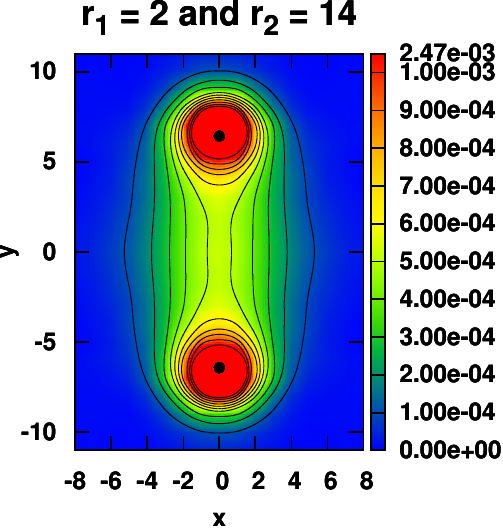}
    \includegraphics[width=4.2cm]{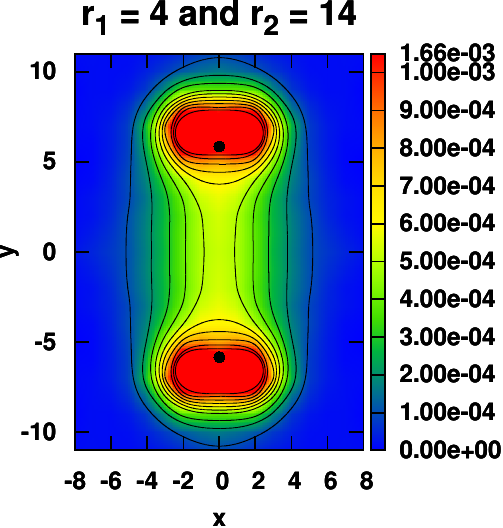}
    \includegraphics[width=4.2cm]{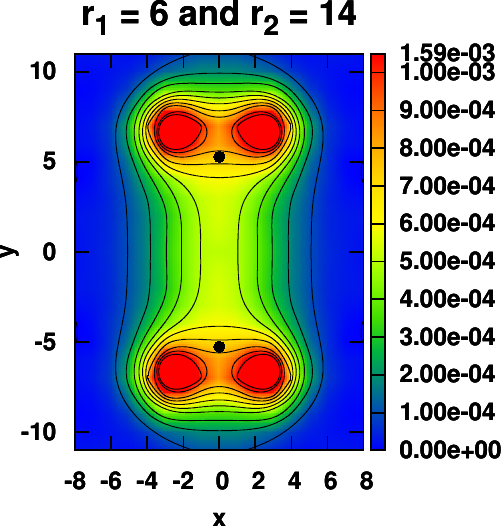}
    \caption{Lagrangian density for $r_2=14$ and $r_1$ from 0 to 6. The black dot points correspond to the Fermat points. 
The results are presented in lattice spacing units (colour online).}
    \label{Act_ape_hyp_r2_14}
\end{center}
\end{figure*}

\begin{figure*}
\begin{centering}
    \subfloat[$\Braket{E^2}$\label{fig:TQ_EB_ape_hyp_r1_8_r2_14_E_Sim}]{
\begin{centering}
    \includegraphics[width=4.2cm]{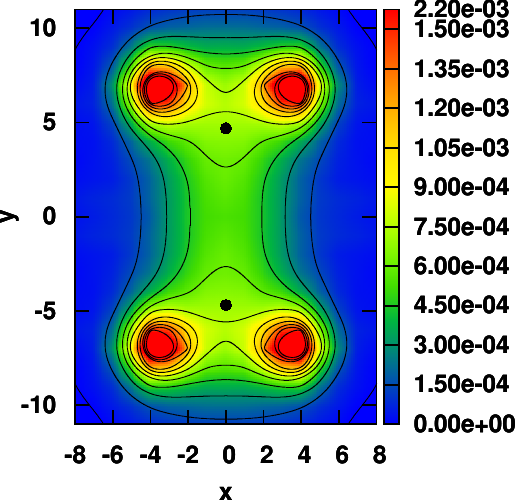}
\par\end{centering}}
    \subfloat[$-\Braket{B^2}$\label{fig:TQ_EB_ape_hyp_r1_8_r2_14_B_Sim}]{
\begin{centering}
    \includegraphics[width=4.2cm]{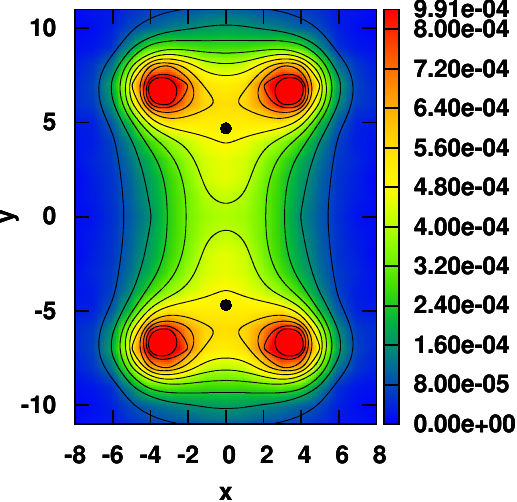}
\par\end{centering}}
    \subfloat[Energy Density\label{fig:TQ_EB_ape_hyp_r1_8_r2_14_Energ_Sim}]{
\begin{centering}
    \includegraphics[width=4.2cm]{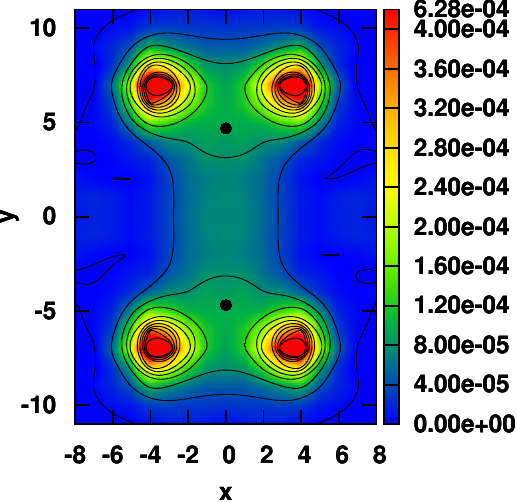}
\par\end{centering}}
    \subfloat[Lagrangian Density\label{fig:TQ_EB_ape_hyp_r1_8_r2_14_Act_Sim}]{
\begin{centering}
    \includegraphics[width=4.2cm]{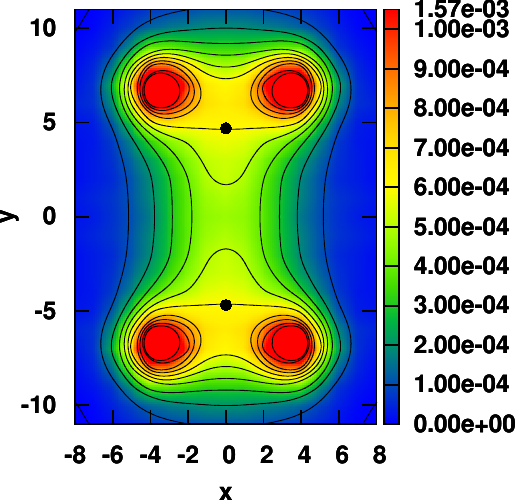}
\par\end{centering}}
\par\end{centering}
    \caption{Colour fields, energy density and Lagrangian density for $r_1=8$ and $r_2=14$. The black dot points correspond to the Fermat points. 
The results are presented in lattice spacing units (colour online).}
    \label{ape_hyp_r1_8_r2_14}
\end{figure*}

To produce the results presented in this work , we use 1121 quenched configurations in a $24^3 \times 48$ lattice at $\beta = 6.2$.
To test whether these configurations are already close to the continuum limit,
we first compare the quark-antiquark static potential obtained using these configurations 
with the potential of 381 configurations in a larger, $32^3\times 64$ lattice, at the same $\beta$. 
The resulting quark-antiquark static potentials are identical within the statistical error, showing that the volume size effects are sufficiently small in our $24^3 \times 48$ lattice.
We present our results in lattice spacing units of $a$, with $a=0.07261(85)$ fm or $a^{-1}=2718\,\pm\, 32$ MeV. 
We generate our configurations in NVIDIA GPUs of the FERMI series (480, 580 and Tesla 2070) with a SU(3) CUDA code
upgraded from our SU(2) combination of Cabibbo-Marinari
pseudoheatbath and over-relaxation algorithm \cite{Cardoso:2010di,ptqcd}.
Our SU(3) updates involve three SU(2) subgroups, we work with 9 complex numbers,
and we reunitarize the matrix.
We have two options to save the configurations, either in a structure of arrays where
each array lists a given complex component for all the lattice sites, or
in an array of structures where each structure is a SU(3) matrix.

In our simulations, the quarks are fixed at $(\pm\,r_1/2,-r2/2,0)$ and the antiquarks at $(\pm\,r_1/2,r_2/2,0)$, with $r_1$ extending up to 8 lattice spacing units and $r_2$ 
extended up to 14 lattice spacing units, in order to include the relevant cases where $r_2 > \sqrt 3 r_1$. 
Notice that in the string picture, at the line
 $r_2 = \sqrt 3 r_1$  in our $(r_1, \, r_2)$ parameter space, the transition between the 
double-Y, or butterfly, tetraquark geometry in Fig. \ref{fig:tq1} to the meson-meson geometry  should occur.
The results are presented only for the $xy$ plane since the quarks are in this plane and the results with $z\neq 0$ are less interesting for this study.
The flux tube fields can be seen in Fig. \ref{Act_ape_hyp_r1_8_r2_14}, \ref{Act_ape_hyp_r2_14} and \ref{ape_hyp_r1_8_r2_14}. 
Theses figures exhibit clearly tetraquark double-Y, or butterfly, shaped flux tubes. The
flux tubes have a finite width, and are not infinitely thin as in the string models inspiring the Fermat points and the triple flip-flop potential, but nevertheless the junctions are close to the Fermat points, thus justifying the use of string models for the quark confinement in constituent quark models.

\begin{figure}[t!]
\begin{center}
    \includegraphics[width=8.5cm]{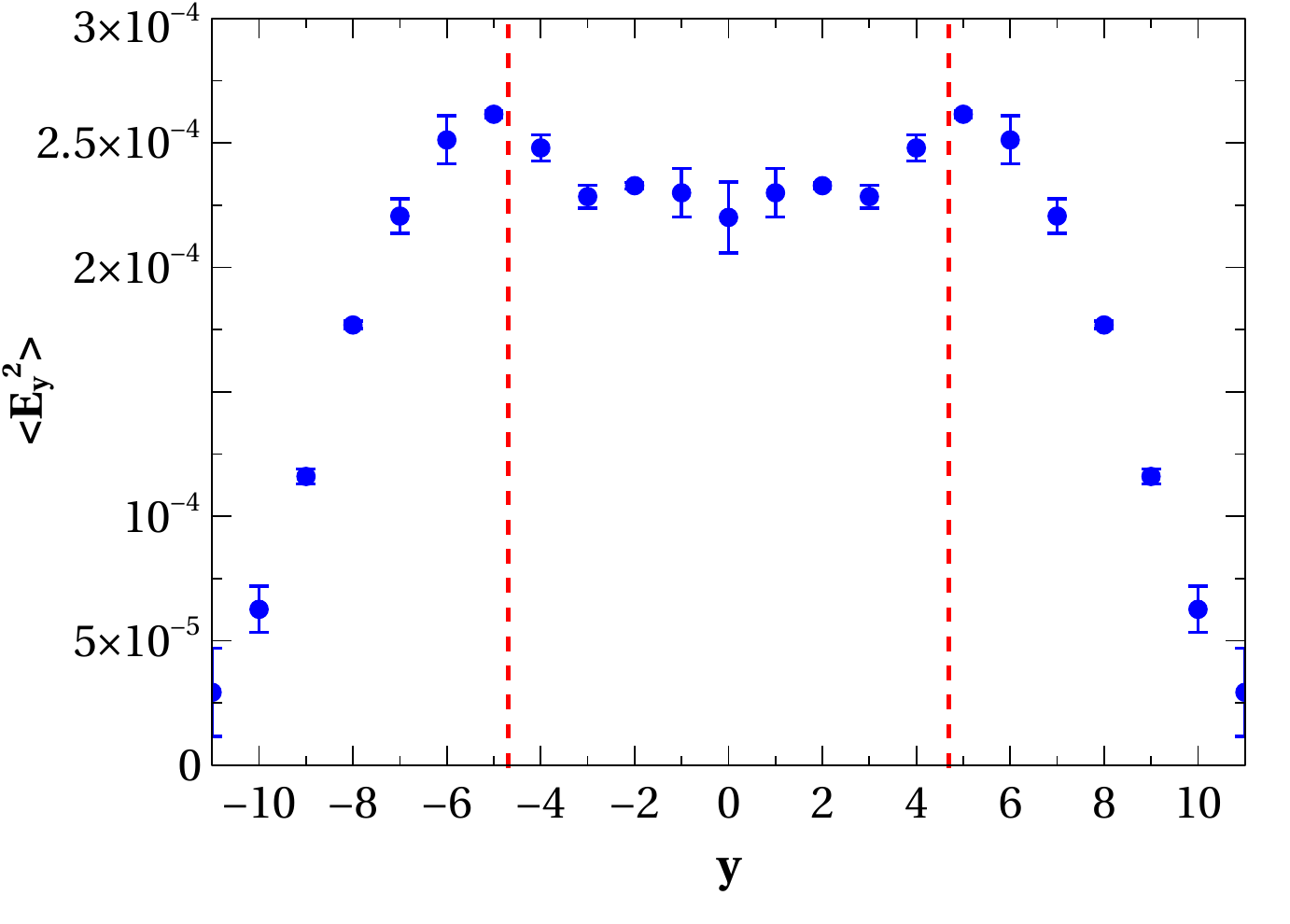}
    \caption{$\Braket{E_y^2}$ in the central axis  $x=0$ for $r_1=8, \ r_2=14$. We show with vertical dashed lines the location of the two Fermat points. The results are presented in lattice spacing units (colour online).}
    \label{fig:Ey_x0}
\end{center}
\end{figure}

In Fig. \ref{fig:Ey_x0}, we plot the chromoelectric field along the central flux tube, $\Braket{E_y^2}$ at $x=0$, for $r_1=8, \, r_2=14$. As expected, the chromoelectric field along $y$ is in agreement with the position of the Fermat points. The chromoelectric field along the $x=0$ central axis is maximal close to the Fermat points situated
at $x \simeq -4.69$ and at $x \simeq 4.69$, flattens in the middle of the flux tube.
Outside the flux tube,  the chromoelectric field is almost residual.

\begin{figure}[t!]
\begin{center}
    \includegraphics[width=7.5cm]{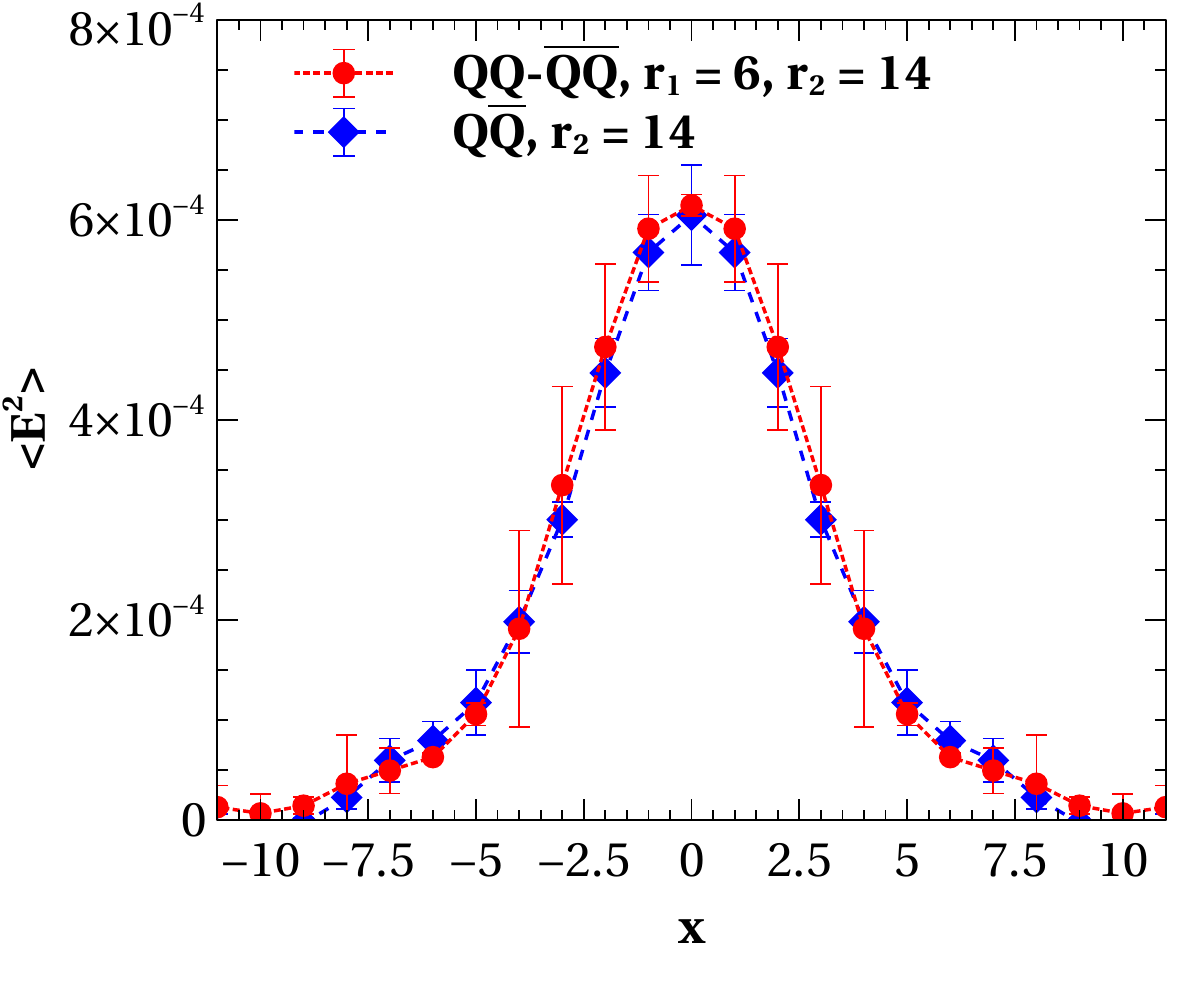}
    \caption{Profile cut at $y=0$ of the chromoelectric field for the tetraquark and quark-antiquark systems in the middle of the flux tube. The results are presented in lattice spacing units (colour online).}
    \label{fig:TQ_QQ_profile}
\end{center}
\end{figure}

In Fig. \ref{fig:TQ_QQ_profile}, we compare the chromoelectric field for the tetraquark and the quark-antiquark system in the middle of the flux tube between the (di)quark and the (di)antiquark.
As can be seen, for our larger distance $r_2=14$ where the source effects are small,
the chromoelectric field is identical up to the error bars, and this confirms that the tetraquark 
flux tube is composed of  a set of fundamental flux tubes with Fermat junctions.

To check which of the colour structures, tetraquark or meson-meson, produces the groundstate flux tube, we study the $\chi^2 /$dof of the $T$ plateaux. Clearly, as expected, the X-shaped geometry of Fig. \ref{fig:tq0} never produces acceptable  plateaux in the range where the meson-meson plateaux are good. 
But, surprisingly, event at distances as small as $r_2 \simeq {1 \over 2}  r_1 \sqrt 3$,  
illustrated in Fig. \ref{fig:TQ_EB_ape_hyp_r1_8_r2_8_Act_3D_Sim}, where the flip-flop potential favours the two-meson flux tube, we find $T$ plateaux with a good   $\chi^2$ /dof. This shows that the mixing between the tetraquark flux tube and the meson-meson flux tube is small, and it is possible to study clear tetraquark flux tubes even at relatively small quark-antiquark distances.

\begin{acknowledgments}
This work was partly funded by the FCT contracts, PTDC/FIS/100968/2008,  CERN/FP/109327/2009 and CERN/FP/116383/2010.
Nuno Cardoso is also supported by FCT under the contract SFRH/BD/44416/2008.
\end{acknowledgments}

\bibliographystyle{apsrev4-1}
\bibliography{bib}

\end{document}